# Selecting the Best Traffic Scheme for the Bicutan Roundabout: A Microsimulation Approach with Multiple Driver-Agents

Merly F. Tataro,[1] Marian G. Arada,[2] and Jaderick P. Pabico[3]

[1,2]Polytechnic University of the Philippines – Taguig, Taguig City 1632, Metro Manila
[3]Institute of Computer Science, University of the Philippines Los Baños, College 4031, Laguna

[1]mftataro@pup.edu.ph, [2]mgarada@pup.edu.ph, [3]jppabico@up.edu.ph

**Abstract** – *Understanding the dynamics of vehicular traffic is important to planning and improving shared public roadways, not only to effectively and efficiently facilitate the controlled movement of vehicles, but also to guarantee the safety of individuals in the vehicles. Because data from real-world traffic is difficult to measure, and conducting replicated experiments on humans has ethical and economical questions, selecting the best scheme out of the proposed vehicular traffic controls has been confined only through simulation.*

*We present the result of our microsimulation study on the effects of six traffic schemes $T=\{t_0, t_1, ..., t_5\}$ on the mean total delay time ($\Delta$) and mean speed ($\Sigma$) of vehicles at the non-signalized Bicutan Roundabout (BR) in Upper Bicutan, Taguig City, Metro Manila, with $t_0$ as the current traffic scheme being enforced and $t_{i>0}$ as the proposed ones. We present first that our simulation approach, a hybridized multi-agent system (MAS) with the car-following and lane-changing models (CLM), can mimic the current observed traffic scenario C at a statistical significance of $\alpha=0.05$. That is, the respective absolute differences of the $\Delta$ and $\Sigma$ between C and $t_0$ are not statistically different from zero. Next, using our MAS-CLM, we simulated all proposed $t_{i>0}$ and compared their respective $\Delta$ and $\Sigma$. We found out using DMRT that the best traffic scheme is $t_3$ (i.e., when we converted the bi-directional 4-lane PNR-PNCC road into a bi-directional 1-lane PNR-to-PNCC and 3-lane PNCC-to-PNR routes during rush hours). Then, we experimented on converting BR into a signalized junction and re-implemented all $t_3$ with controlled stops of $S=\{15s, 45s\}$. We found out that $t_3$ with a 15-s stop has the best performance. Finally, we simulated the effect of increased in vehicular volume V due to traffic improvement and we found out that $t_3$ with 15-s stop still outperforms the others for all increased in $V=\{10\%, 50\%, 100\%\}$.*

## I. INTRODUCTION

Transportation is an important aspect of both the economy and the lifestyle of the population it is serving because it links the various destinations of people, goods, and services. A smoother and faster vehicular and pedestrian traffic would mean people reaching their workplaces on time resulting into higher number of productive hours. Increased productivity results into higher wages and profit which in return encourage people to buy more vehicles. The problem of traffic congestion in our country is the result mostly of having more automotive growth than the available road can accommodate. According to the Metro Manila Development Authority's (MMDA) report recently, vehicular speed drops to 5kph during rush hours along Epifanio Delos Santos Avenue (EDSA). The MMDA also reports that major roads are no longer sufficient to accommodate the consistently rising traffic volume, which is made worse by the continuing use of outmoded traffic signals, worsening road conditions, and decreasing availability of efficient public transport [1]. As a result, mere accident at an important intersection, for example, is likely to paralyze many streets for hours [2]. To address this seemingly worsening problem, especially in the Metro Manila area, the government has implemented various infrastructure projects including widening of roads, and construction of overpasses and underpasses along major thoroughfares. One among the many bottleneck areas in Metro Manila that is being considered for infrastructure improvement is the Bicutan Roundabout (BR) in Taguig City. Both the MMDA and the City Government of Taguig (CGT) are separately considering building infrastructure to improve the traffic condition in the area [1,3].

BR is located at the Upper Bicutan area and has an inscribed diameter of 34m. It is bounded to the East by the Department of Science and Technology (DOST) campus, to the North by the Philippine National Construction Corporation (PNCC) campus, to the West by the Philippine National Railways (PNR) crossing, and to the South by North Daanghari. BR serves as the T-intersection between General Paulino Santos Avenue (GPSA) that lies along the ENE-WSW line and SLEX's East Service Road (ESR) that lies along the NEN-SWS line of the area (Figure 1). The nominal center of BR is approximately 100m WSW from DOST's main gate along GPSA, 150m SWS of the ESR and San Martin de Porres fork, and 106m ENE of PNR crossing. Jeepneys, buses, trucks, taxis, AUVs, motorcycles, tricycles, and bicycles are the usual vehicles that pass the area.

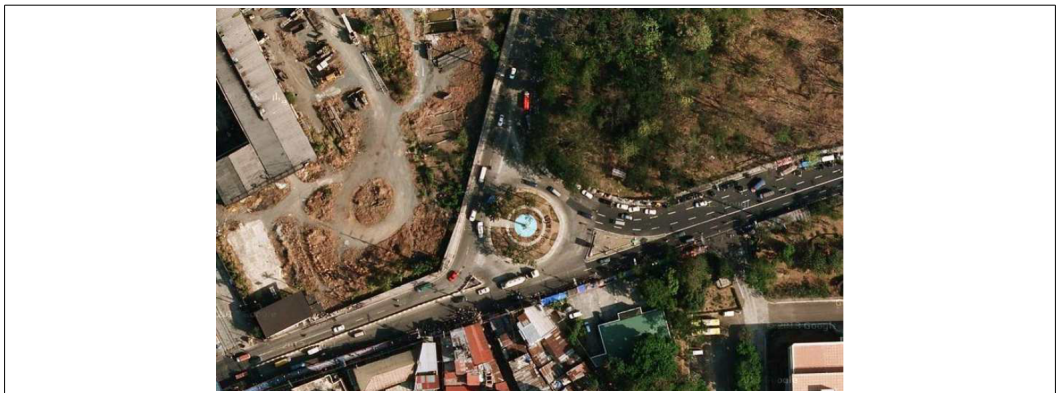

**Figure 1.** The area map of Upper Bicutan, Taguig City showing BR.

Because BR is a bottleneck, it has become common to see and experience traffic congestion in the area especially at peak hours both in the morning and in the afternoon. Even though BR is non-signalized, the local government unit has to employ traffic enforcers to manually intervene with controlling the traffic flow during these hours to ease congestion in the area [3]. The enforcers also provide some sort of relief to commuters when they see that city official are at least doing something [4]. The traffic enforcers react to the traffic situation by implementing *ad hoc* solutions that only solve the various current situations, instead of a programmed solution that can solve the big problem in the long run [5,6]. This problem is expected to continue to worsen in the future due to the rapid growth of population that is aggravated further by the increasing number of vehicles passing in the area [1].

In an earlier works by Arada, et al. [5,6], they simulated the effect of various proposed infrastructure developments in BR on the mean total vehicular delay time ($\Delta$) and on the mean speed ($\Sigma$). They found out that the best infrastructure development ($ID_{best}$) that minimizes $\Delta$ and maximizes $\Sigma$, out of the three that are being considered, is the one that widens the eastbound lane of the GPSA from the PNR towards the DOST gate. They were even successful in finding out the effect to $\Delta$ and $\Sigma$ of increasing the vehicular volume up to 100% in the BR area under the $ID_{best}$ scenario. However, before we support to recommend the implementation of the findings of Arada, et al. [5,6], we caution the reader that there may exists a traffic scheme that can approximate the benefits of $ID_{best}$ in terms of $\Delta$ and $\Sigma$. If we can find this, the government is posed to save a lot in construction costs and time. Thus, in this research effort, we state the following hypotheses:

**H$_0$**: (<u>Null Hypothesis</u>) Among the schemes that are available, there does not exist a traffic scheme that can at least approximate the optimal $\Delta$ and $\Sigma$ that the $ID_{best}$ can provide.

**H$_1$**: (<u>Alternate Hypothesis</u>) Among the schemes that are available, there does exist a traffic scheme that can at least approximate the optimal $\Delta$ and $\Sigma$ that the $ID_{best}$ can provide.

To test these hypotheses, we used the methodology by Arada, et al. [5,6] in employing a hybridized microsimulation with a multi-agent system (MAS) and the car-following and lane-changing models (CLM) on the effect of six traffic schemes $T=\{t_0, t_1, ..., t_5\}$ on the mean Δ and mean Σ. The $t_0$ is the current *ad hoc* scheme being implemented while $t_{i>0}$ are the ones with suggested long-term programming. Because data from real-world traffic is difficult to measure, and conducting replicated experiments on humans has ethical and economical questions, selecting the best scheme out of the proposed vehicular traffic controls can only be realistically done through simulation. Thus, first, we show statistically that our MAS-CLM can simulate the real-world scenario, just like how Arada, et al. [5,6] did it. Second, we use MAS-CLM to simulate various traffic schemes and compile their respective mean Δ and mean Σ. Lastly, we simulate the effect of increased vehicular volume on Δ and Σ under the best traffic scheme condition.

## II. THE MAS-CLM MODEL

Our MAS-CLM model, described in detail in the works of Arada, et al. [5,6] and of Tataro, et al. [4], uses a car-following model developed by Herman and Rothery [7]. Relating this model to our MAS, a driver-agent is said to be in the following state when it is constrained by the speed of the moving vehicles that is directly in front of it. Speeding at the driver-agent's desired speed may lead to collision under this state. However, when the driver-agent is not constrained by another vehicle, then it is said to be in the free state and travels, on the average, at its desired speed. The action of the driver-agent while in the following state is defined by the driver-agent's acceleration, although in some models like the one developed by Gipps [8] the action is dependent on the driver-agent's speed. Our MAS-CLM model uses the following states to describe the driver-agent's behavior (the pertinent details of this model, including the mathematical foundations of the equation of motion has already been discussed elsewhere (see Arada, et al. [5,6]):

1. <u>Free Driving State</u> – The driver-agent at this state is unconstrained and tries to achieve its goal speed $v_{goal}$, subject of course to pertinent legal speed limits being imposed at the portion of the road. If the driver-agent's current speed $v_{current} > v_{goal}$, the driver-agent uses the normal deceleration rate $-a_{norm}$ to slow down to $v_{goal}$. If $v_{current} < v_{goal}$, the driver-agent uses the maximum acceleration rate $a_{max}$ to speed up to $v_{goal}$ at the shortest possible time $t_{min} = v_{goal}/a_{max}$. The $a_{norm}$ and $a_{max}$ are parameters of the driver-agents and dependent on the vehicular type being simulated.
2. <u>Normal Following State</u> – The driver-agent under this state adjusts its speed with respect to the speed of the vehicle directly in front of it. Its acceleration is defined by the asymmetrical Gazis-Herman-Rothery (GHR) model described by Chandler, et al. [9], Gazis, et al. [10], and Yang and Koutsopoulos [11].
3. <u>Emergency Deceleration State</u> – The driver-agent while at this state tries to decelerate to avoid colliding with the slow-moving vehicle directly in front of it.

## III. METHODOLOGY

We present in this section the different activities we performed to test the hypotheses that we stated above. We also present here the metrics that we used and the statistical analyses that we employed to analyze our results.

*A. Observation of the current traffic condition*

We assigned twelve (12) research assistants to different identified points within the BR to record the time of entrance and exit of the vehicle, and the type of vehicle that passes through. They were tasked to record the plate numbers of the vehicles as well as to take videos of the same vehicles that pass through certain points of entrance and exit within the BR. We conducted a series of actual observations during peak and off peak hours to get an accurate data and determine the highest and the lowest traffic volume. From these observations, we obtained the mean travel time of sampled vehicles, as well as the respective distributions of each vehicle type. The following are the points of entrances and exits that we designated as observation points:
1. Point A (PNR) is the route from PNR passing through the BR to Point B or C;
2. Point B (PNCC) is the route from PNCC passing through the BR to Point A or C; and
3. Point C (DOST) is the route from DOST passing through the BR to Point A or B.

*B. Simulation of the current traffic condition*

We conducted a replicated microsimulation of the vehicular flow under the current BR using the data on respective distributions of vehicles by type. We replicated the study to $n = 10$ and computed $\Delta$ and $\Sigma$. We wanted to know if the simulated data $t_0$ can statistically reproduce the current condition $C$, under the same traffic scheme. Statistically, the respective differences of $\Delta$ and $\Sigma$ between $C$ and $t_0$ must not be different from zero at $\alpha = 0.05$. We used analysis of variance statistics (ANOVA) to evaluate two hypotheses as follows:

**H$_2$**: (Null Hypothesis) There is no significant difference between $C$ and $t_0$ at $\alpha = 0.05$, i.e., $|C - t_0| \equiv 0$, where the symbol $\equiv$ means statistically equal at 5% confidence level.

**H$_3$**: (Alternate Hypothesis) There is significant difference between $C$ and $t_0$ at the same $\alpha$, i.e., at 5% confidence, $|C - t_0| > 0$.

An example visualization of one our microsimulation runs is shown in Figure 2. The visualization uses the VisSim framework for transport planning, traffic engineering and traffic simulation [12].

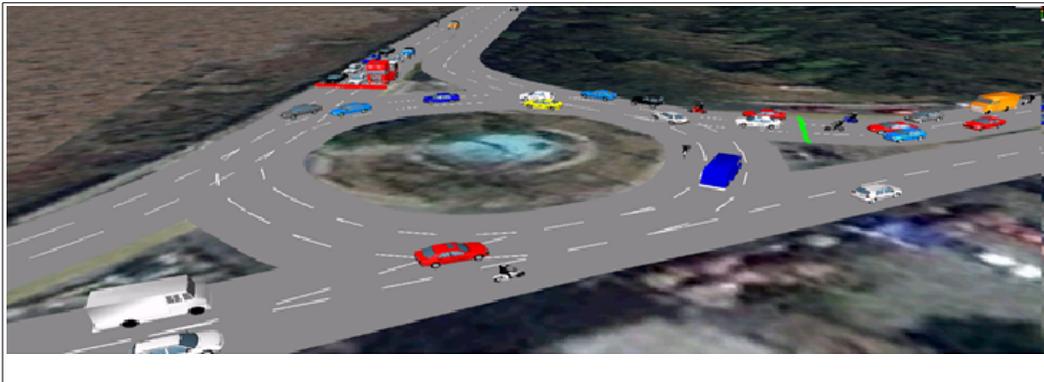

**Figure 2.** A screenshot of an example microsimulation run on BR using VisSim as a simulation visualization framework. The red line across the road means that the vehicles coming from that direction are stopped, while the green line means that the vehicles are moving.

*C. Microsimulation of the proposed traffic schemes*

We created the respective replicated $n = 10$ microsimulations of the vehicular flow using the vehicular distribution in $C$ under the five proposed traffic schemes $t_1, t_2, ..., t_5$ as follows:

$t_1$: Controlled one-way scheme for short-timed intervals of 60s time cycle: 30s stop those coming from Point B followed by 30s stop those coming from Point C;

$t_2$: Same as $t_1$ but the time interval is 120s;

$t_3$: Designation of one lane from Point A to Point B and three lanes in the opposite lane (from Point B to Point A) during peak hours instead of the normal two-lane, two-way traffic scheme;

$t_4$: Combination of $t_1$ and $t_3$; and

$t_5$: Combination of $t_2$ and $t_3$.

We computed the respective means of $\Delta$ and $\Sigma$ for all proposed $t_1, t_2, ..., t_5$, including that of the $t_0$ and conducted ANOVA to find out whether the respective means are significantly different from each other at $\alpha = 0.05$. We, thus, state the following hypotheses (where $\Delta(t_i)$ means the $\Delta$ under $t_i$):

**H$_4$**: (Null Hypothesis) The absolute pairwise differences $|\Delta(t_0)-\Delta(t_1)|\equiv\cdots\equiv|\Delta(t_1)-\Delta(t_2)|\equiv\cdots\equiv|\Delta(t_2)-\Delta(t_3)|\equiv\cdots\equiv|\Delta(t_4)-\Delta(t_5)|\equiv 0$.

**H$_5$**: (Alternate Hypothesis) At least one of the following respective pairwise absolute differences is true: $|\Delta(t_0)-\Delta(t_1)|>0$, ..., $|\Delta(t_1)-\Delta(t_2)|>0$, ..., $|\Delta(t_2)-\Delta(t_3)|>0$, ..., or $|\Delta(t_4)-\Delta(t_5)|>0$.

We have a similar null **H$_6$** and alternate **H$_7$** hypotheses for $\Sigma$, where $\Sigma(t_i)$ means the $\Sigma$ under $t_i$:

**H$_6$**: (Null Hypothesis) The absolute pairwise differences $|\Sigma(t_0)-\Sigma(t_1)|\equiv\cdots\equiv|\Sigma(t_1)-\Sigma(t_2)|\equiv\cdots\equiv|\Sigma(t_2)-\Sigma(t_3)|\equiv\cdots\equiv|\Sigma(t_4)-\Sigma(t_5)|\equiv 0$.

**H$_7$**: (Alternate Hypothesis) At least one of the following respective pairwise absolute differences is true: $|\Sigma(t_0)-\Sigma(t_1)|>0$, ..., $|\Sigma(t_1)-\Sigma(t_2)|>0$, ..., $|\Sigma(t_2)-\Sigma(t_3)|>0$, ..., or $|\Sigma(t_4)-\Sigma(t_5)|>0$.

*D. Signalized BR with controlled stops and increased volume*

We experimented on converting BR into a signalized junction and re-implemented the best scheme found above with controlled stops of 15s and 45s, making sure that the distribution by vehicular type is preserved. Here, we wanted to know if the benefits of the best proposed scheme will be carried with controlled signal stops, and if so, up to how much increase? We then reran the respective replicated microsimulations with respective increases of 10%, 50%, and 100% in vehicular volume $V$, again making sure that the distribution by vehicular type is preserved. We wanted to know if the benefits of the proposed schemes will still be carried with the increase in $V$ and if so, up to how much increase within the 0% to 100% range? The assumed increase in $V$ is just a natural reaction of the driving agents when there is a perceive improvement in the current situation. The improvement of the vehicular flow due to improved traffic schemes will attract more driver-agents, increasing $V$ in the area, and thereby might degrade the expected designed benefits of the schemes in the long run.

## IV. RESULTS AND DISCUSSION

*A. Observed vs. Simulated Current Traffic Condition*

Table 1 summarizes the statistics we computed after conducting ANOVA on $C$ and $t_0$. We see that the F statistics $\alpha_F = 0.6873 > \alpha = 0.05$, which prompts us to accept **H$_2$** and say that the difference between the mean sampled observed data $C$ and the mean simulated data $t_0$ is not significantly different from zero at $\alpha \approx 0.05$. We can, thus, say that our microsimulation model was able to mimic the driver's behavior in the real world with a guarantee of being correct $1 - \alpha = 0.95$ of the time.

**Table 1.** ANOVA table comparing $C$ and $t_0$ using the F statistics. SOV means sources of variations, and DF means degrees of freedom.

| SOV | DF | Sum of Squares | Mean Square | F | $\alpha_F$ |
| --- | --- | --- | --- | --- | --- |
| Replication | 9 | 1,633.05 | 181.45 | 39.21 | < 0.0001 |
| $C$ vs. $t_0$ | 1 | 0.80 | 0.80 | 0.17 | 0.6873 |
| Error | 9 | 41.65 | 4.63 | | |
| Total | 19 | 1,675.50 | | | |

*B. Comparison of the Traffic Schemes*

Tables 2 and 3 show the respective ANOVA of $\Delta$ and $\Sigma$. We find that $\alpha_F(\Delta) < 0.0001$ and $\alpha_F(\Sigma) < 0.0001$, which prompts us to accept **H$_5$** and **H$_7$** for $\Delta$ and $\Sigma$, respectively. This means that at least one of the absolute pairwise differences between $\Delta(t_i)$ and $\Delta(t_j)$, and those between $\Sigma(t_i)$ and $\Sigma(t_j)$, $\forall i \neq j$, are significantly greater than zero for $\Delta$ and $\Sigma$, respectively.

**Table 2.** ANOVA table comparing Δ at $T$ (i.e., $\Delta(t_0)$ vs. $\Delta(t_1)$ vs. ... vs. $\Delta(t_5)$).

| SOV | DF | Sum of Squares | Mean Square | F | $\alpha_F$ |
|---|---|---|---|---|---|
| $T$ | 5 | 37,354.14 | 7,470.83 | 319.01 | < 0.0001 |
| Error | 54 | 1,264.61 | 23.42 | | |
| Total | 59 | 38,618.75 | | | |

Figure 3 shows the pairwise comparison of the mean Δ and the mean Σ under various $T=\{t_0, t_1, ..., t_5\}$ using the Duncan's Multiple Range Test (DMRT). We can see from the table that the best traffic scheme is $t_3$, both in terms of Δ and Σ. From the figure, we can also see the relative mean Δ and mean Σ of $ID_{best}$ found by Arada, et al. [5,6]. Even though the data for $ID_{best}$ are not included in our DMRT, we can see relatively that even both $\Delta(t_{i>0})$ and $\Sigma(t_{i>0})$ outperform $ID_{best}$. We, thus, accept **H**$_1$ and say in fact that all traffic schemes that we considered here outperform $ID_{best}$ at both Δ and Σ.

**Table 3.** ANOVA table comparing Σ at $T$ (i.e., $\Sigma(t_0)$ vs. $\Sigma(t_1)$ vs. ... vs. $\Sigma(t_5)$).

| SOV | DF | Sum of Squares | Mean Square | F | $\alpha_F$ |
|---|---|---|---|---|---|
| $T$ | 5 | 917.46 | 183.49 | 483.58 | < 0.0001 |
| Error | 54 | 20.49 | 0.38 | | |
| Total | 59 | 937.95 | | | |

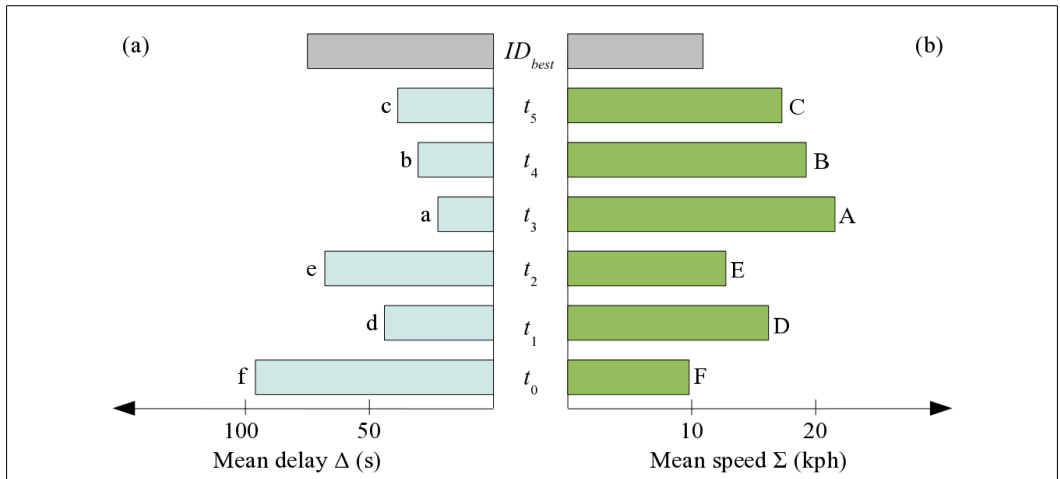

**Figure 3.** Pairwise comparison of (a) mean Δ and (b) mean Σ over various traffic schemes $T=\{t_0, t_1, ..., t_5\}$. Means with the same small letters (a) or the same capital letters (b) are not significantly different by DMRT at α=0.05. $ID_{best}$ is the mean Δ and mean Σ of the best infrastructure development found by Arada, et al. [5,6].

*C. Effect of signalized $t_3$*

When we converted BR into a signalized junction and reimplemented $t_3$ with controlled stops of 15s and 45s (i.e., $t_3$+15s and $t_3$+45s), respectively, we found out that the metrics Δ and Σ improved statistically over the non-signalized scheme (Figure 4). Specifically, $\Delta(t_3$+15s$)$ is statistically better than both $\Delta(t_3)$ and $\Delta(t_3$+45s$)$. Consequently, $\Delta(t_3)$ is statistically better than $\Delta(t_3$+45s$)$. This means that among the three schemes, $\Delta$ is at minimum under $t_3+15s$. Additionally, $\Sigma(t_3)$ and $\Sigma(t_3+15s)$ are statistically the same, as well as $\Sigma(t_3)$ and $\Sigma(t_3+45s)$. However, $\Sigma(t_3+15s)$ is statistically better than $\Sigma(t_3+45s)$. This means that $\Sigma$ is at maximum under either $t_3$ or $t_3+15s$. Since the best $\Delta$ is under $t_3+15s$ and the best $\Sigma$ is under either $t_3$ or $t_3+15s$, we recommend that $t_3+15s$ be implemented in BR. Thus, we say that installing signal lights with 15s controlled stops within the scheme described by $t_3$ will provide statistically significant improvement to the traffic flow.

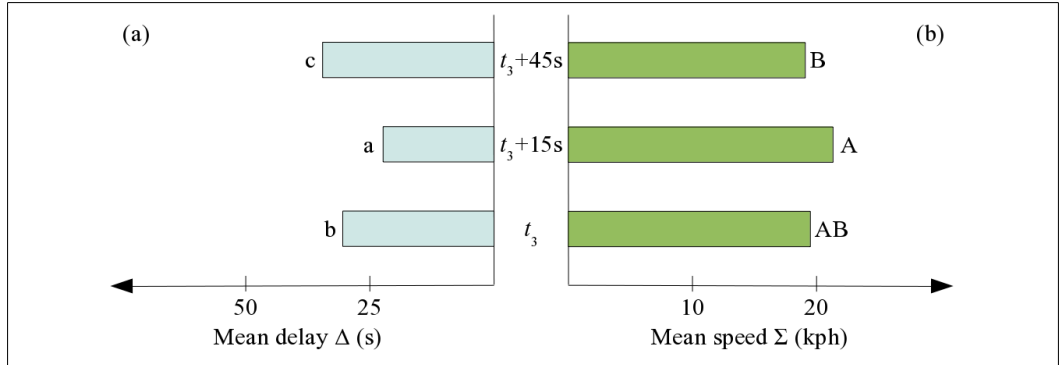

**Figure 4.** Pairwise comparison of (a) mean $\Delta$ and (b) mean $\Sigma$ over the traffic schemes $t_3$, $t_3+15s$, and $t_3+45s$. Means with the same small letters (a) or the same capital letters (b) are not significantly different by DMRT at $\alpha=0.05$.

*D. Effect of increased V to a signalized $t_3$*

Figure 5 shows the effect to the metrics $\Delta$ and $\Sigma$ of increasing $V$ when the traffic scheme $t_3+15s$ is implemented. The respective increases are 10%, 50%, and 100% of the observed volume. The $\Delta$ increased when $V$ was increased (Figure 5a), while $\Sigma$ decreased when $V$ was increased (Figure 5b). Increasing $V$ to 10% increased $\Delta$ significantly for $t_3+15s$, but the increase is still not enough to be greater than either $t_3$ or $t_3+45s$ (Figure 5a inset). At additional $V$ of 10%, $t_3+15s$ is still a better scheme than $t_3$, while $t_3$ is better than $t_3+45s$. Equations 1 through 6 show the respective linear estimates of $\Delta(t_3)$, $\Delta(t_3+15s)$, $\Delta(t_3+45s)$, $\Sigma(t_3)$, $\Sigma(t_3+15s)$, and $\Sigma(t_3+45s)$, as $V$ is increased up to 100%. In these equations, $V^{(+)}$ is the percent increase in $V$, $r^2$ is the coefficient of determination, and an asterisk and an "ns" over a coefficient respectively means that the coefficient is significantly and not significantly different from zero via a two-tailed $t$-test.

$$\Delta(t_3) = 4.58^* \, V^{(+)} - 2.69^{ns}, \quad r^2 = 0.97 \tag{1}$$
$$\Delta(t_3+15s) = 0.47^* \, V^{(+)} + 21.92^*, \quad r^2 = 0.94 \tag{2}$$
$$\Delta(t_3+45s) = 0.91^* \, V^{(+)} + 34.82^{ns}, \quad r^2 = 0.90 \tag{3}$$
$$\Sigma(t_3) = -0.15^* \, V^{(+)} + 18.43^*, \quad r^2 = 0.95 \tag{4}$$
$$\Sigma(t_3+15s) = -0.08^* \, V^{(+)} + 21.02^*, \quad r^2 = 0.96 \tag{5}$$
$$\Sigma(t_3+45s) = -0.11^* \, V^{(+)} + 18.47^{ns}, \quad r^2 = 0.85 \tag{6}$$

Since all slope coefficients are significantly different from zero, we can say that further increasing $V^{(+)}$ to 50% and to 100% increased $\Delta(t_3)$, $\Delta(t_3+15s)$, and $\Delta(t_3+45s)$ linearly by 4.58s/% (Equation 1), 0.47s/% (Equation 2), and 0.91s/% (Equation 3), respectively. It is clear to see that up to an increased $V$ of 100%, $t_3+15s$ is the best scheme with respect to minimizing the rate of increase for $\Delta$.

In Figure 5b, increasing the additional $V$ up to $V^{(+)}=100\%$ linearly decreases $\Sigma$. The linear decrease for $\Sigma(t_3)$, $\Sigma(t_3+15s)$, and $\Sigma(t_3+45s)$ are, respectively, 0.15kph/% (Equation 4), 0.08kph/% (Equation 5), and 0.11kph/% (Equation 6). With respect to minimizing the rate of decrease for $\Sigma$, the best traffic scheme is $t_3+15s$. Thus, it is recommended that a signalized version of $t_3$ with stops of 15s be implemented in BR.

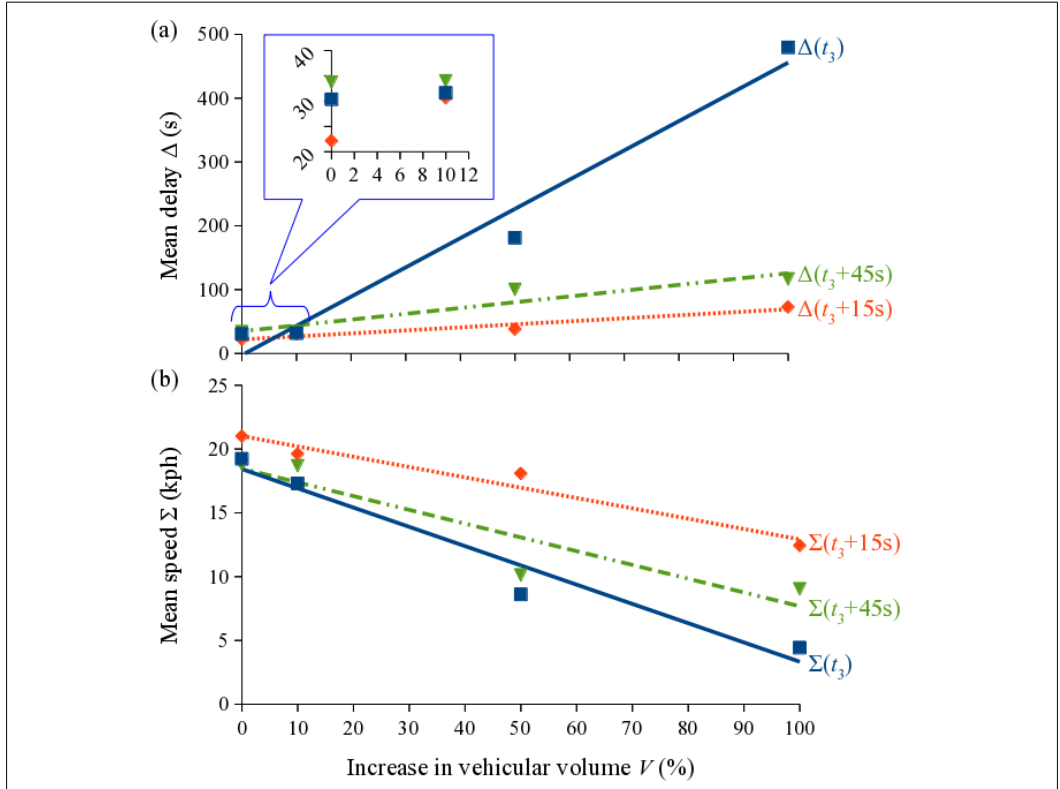

**Figure 5.** Effect on (a) mean $\Delta$ and (b) mean $\Sigma$ on increasing $V$ to 10%, 50%, and 100%. The markers are mean data while the lines are the regression lines defined in Equations 1 through 6.

V. SUMMARY, CONCLUSION, RECOMMENDATION AND EXTENSION

In this paper, we presented our MAS-CLM which we used to simulate six traffic schemes in the nonsignalized BR. We found out that our MAS-CLM can statistically simulate the real-world at $\alpha=0.05$. Using the same MAS-CLM, we simulated $T=\{t_0, t_1, ..., t_5\}$ and collected $\Delta$ and $\Sigma$ as metrics. DMRT at $\alpha=0.05$ shows that $t_3$ outperforms the other schemes $t_{i\neq3}$. We were also able to show that the traffic schemes $t_{i>0}$ have better mean $\Delta$ and mean $\Sigma$ than the one found in an earlier work involving a costly infrastructure development [5,6]. We converted BR into a signalized roundabout and reimplemented all $t_3$, while we also considered increasing $V$ from 10% up to 100%. We found out that $t_3$ with 15-s stop-cycle outperforms the other two schemes. Because $ID_{best}$ performed poorly compared to $t_{i>0}$, we highly recommend to either MMDA or to CGT to consider implementing $t_3$ with 15-s stop-cycle instead of developing the infrastructure described by $ID_{best}$. We caution the reader, however, that we may be able to find better mean $\Delta$ and mean $\Sigma$ than the one we found in this study if we consider the combination effects of infrastructure development and traffic scheme. Our study on the combination effects are already underway as an extension of this work. We will report the result of this extension in our future paper.

V. ACKNOWLEDGMENTS, DISCLOSURE OF EARLIER PRESENTATION, AND AUTHOR CONTRIBUTIONS

This research endeavor is funded by DOST ASTHRDP Graduate Scholarship Program with MFT and MGA as scholar-grantees taking MIT at UPLB under the research advisorship of JPP. VisSim is courtesy of PTV Planung Transport Verkehr AG of Germany.

An earlier version of this paper was presented as an oral paper [13] during the *6th Annual University of the Philippines Los Baños (UPLB) College of Arts and Sciences (CAS) Student-Faculty Research Conference (SFRC 2013)* held at the NCAS Auditorium, UPLB on 13 December 2013.